# A Conceptual Model for Measuring the Complexity of Spreadsheets

Thomas Reschenhofer, Bernhard Waltl, Klym Shumaiev, Florian Matthes
Technical University of Munich
Munich, Germany
reschenh@in.tum.de, b.waltl@tum.de, klym.shumaiev@tum.de, matthes@in.tum.de

**ABSTRACT**

*Spreadsheets are widely used in industry, even for critical business processes. This implies the need for proper risk assessment in spreadsheets to evaluate the reliability and validity of the spreadsheet's outcome. As related research has shown, the risk of spreadsheet errors is strongly related to the spreadsheet's complexity. Therefore, spreadsheet researchers proposed various metrics for quantifying different aspects of a spreadsheet in order to assess its complexity. However, until now there is no shared understanding of potential complexity drivers for spreadsheets. The present work addresses this research gap by proposing a conceptual model integrating all aspects which are identified by related literature as potential drivers to spreadsheet complexity. In this sense, this model forms the foundation for a structured definition of complexity metrics, and thus enhances the reproducibility of their results. At the same time, it forms the foundation for identifying further applicable complexity metrics from other scientific domains.*

## 1 INTRODUCTION

Spreadsheets are the Swiss Army Knife for decision support in enterprises: They empower business users from different domains to manage, analyze, and visualize their domain-specific data for deriving – very often business-critical – decisions [Panko and Port, 2012; Reschenhofer and Matthes, 2015]. Spreadsheets are used in a variety of application areas, e.g., for financial reporting [Panko, 2006] and workload planning [Pemberton and Robson, 2000]. Due to a wide dissemination of spreadsheets across nearly all business domains [Scaffidi, Shaw and Myers, 2005; Bradley and McDaid, 2009], but their invisibility to the corporate IT departments, Panko and Port [Panko and Port, 2012] call spreadsheets the "dark matter" of IT. However, spreadsheets are not only widely used in a plethora of application areas, but also important and critical for companies [Chan and Storey, 1996; Gable, Yap and Eng, 1991; Hall, 1996]. Therefore, errors in spreadsheets can have significant negative impact [Caulkins, Morrison and Weidemann, 2007; Powell, Baker and Lawson, 2009]. This becomes even more critical since numerous studies [Panko, 2000; Powell, Baker and Lawson, 2008] have shown that spreadsheets are indeed very error-prone. Even worse, users tend to overlook the risk of errors in their spreadsheets since they are not able to assess this risk [Hall, 1996]. Therefore, they often blindly trust the respective outcomes and thus make potentially costly decisions.

Spreadsheets as a generic measure for complex calculation on data are heavily used in the financial industry, e.g., determination of risks based on predefined formulas, measurement of equity requirements using given regulation policies such as Basel II, respectively Basel III, and the Sarbanes-Oxley-Act (SOX). Obviously, spreadsheets have become very popular for banks and insurances [Hall, 1996; Janvrin and Morrison, 2000]. Supervising authorities responsible for auditing, such as the German BaFin (engl. Federal Financial Supervisory Authority), are also aware of the importance of spreadsheets in the financial industry [Bretz, 2012]. Therefore, supervisors have a particular focus on spreadsheets, their structure, and calculation. Based on their experience they also state that complex calculations in spreadsheets are vulnerable to errors and can become an additional source for risks [Bretz, 2009]. The lack of separation between data and logic and the usage of difficult formulas with many dependencies hardens the problem of creation, controlling and maintenance of spreadsheets [Bretz, 2009]. The analysis of spreadsheets regarding complexity based on a conceptual model is a step towards detailed investigation of the structure and semantics of spreadsheets as sources of risks. An important driver for the risk of errors in spreadsheets is the complexity of its design which mainly constituted by the formulas and dependencies between cells [Teo and Lee-Partridge, 2001; Bregar, 2004; Janvrin and Morrison, 2000]. In this sense, the complexity of a spreadsheet correlates with its understandability, and hence is an indicator for the probability of errors. Therefore, determining the complexity helps during the assessment of errors and risks [Bregar, 2004; Hermans, Pinzger and van Deursen, 2012b].

In many domains, metrics are already common to assess the complexity or understandability of artifacts. Analyzing linguistics properties has always been an objective to gain insights to a particular text or discourse or to the structure of written language in general [Graesser and McNamara, 2011]. Several approaches exist aiming at the determination of qualitative and quantitative textual properties [Köhler, 2005]. Linguists have always tried to measure understandability and readability as objective and measurable indices. Those have been used in the assessment of texts for education [Flesch, 1948], journalism, military, healthcare [DuBay, 2004], and recently also in the legal domain [Waltl and Matthes, 2014]. Thereby, they have developed and reused several metrics representing the properties of a text quantitatively. Assessing textual complexity can be achieved by a set of metrics, rather than by one single metric. Consequently, the resulting metrics developed by linguistics can be adapted and reused in other domains in order to investigate and quantify complexity of formulas and dependencies in spreadsheet formulas.

The objective of the present work is to propose a conceptual spreadsheet model forming the foundation for identifying applicable complexity metrics in a structured way, and thus enhances the reproducibility of metric results through a unified spreadsheet model. In this sense, this model captures all aspects of spreadsheets, which are potential drivers for its complexity. Furthermore, this spreadsheet model can serve as a starting point for the identification of metrics from other domains. Based on this model, we refine metrics from the domain of software engineering which were already adapted to spreadsheets, and identify metrics from linguistics, which are also applicable for assessing the complexity of spreadsheets. Finally, we want to answer the question how complex today's spreadsheets are. This leads to the following research questions constituting the present work's contribution:

- What is a spreadsheet model capturing potential complexity drivers for spreadsheets, and which enables the formal definition of complexity metrics?
- How can the metrics from software engineering and linguistics be defined based on the proposed conceptual model?
- According to those indicators, how complex are today's spreadsheets, and how do those metrics correlate to each other?

In order to answer those research questions, the remainder of this paper is organized as follows: Section 2 summarizes related work in the field of spreadsheet research, in particular about risk assessment in spreadsheet. Thereafter, Section 3 describes the applied research methodology. In Section 4 we propose a model for the definition of spreadsheet complexity metrics and thus the answer for the first research question. Based on this model, Section 5 answers the second research question by describing a set of metrics from the domains of software engineering and linguistics, which then are applied to two spreadsheet corpora as shown in Section 6.

## 2 RELATED WORK

Due to the vulnerability of spreadsheets to errors and the potentially high negative impact of those, the assessment of spreadsheets regarding complexity has been subject of recent research. Hermans et al. [Hermans et al., 2012b] correlate the risk and complexity of spreadsheets with the understandability of formulas. They derived a set of metrics for measuring formula understandability by conducting interviews with spreadsheet experts. Those metrics capture both the complexity of formulas and their placement within the spreadsheet. Furthermore, they evaluated their work by correlating the metrics to the perceived understandability by spreadsheet experts. Although Hermans et al. come up with a list of important factors to understand a spreadsheet, they do not provide a formal model based on these findings, which in turn is the main contribution of the present paper. Additionally, while the proposed metrics only assess the understandability at the level of formulas, the present work's spreadsheet model allows the definition of metrics at higher levels, e.g., on the level spreadsheets which can capture the interrelations between formulas and worksheets.

In another publication, Hermans et al. [Hermans, Pinzger and van Deursen, 2014] investigate code smells within spreadsheet formulas. A "code smell" is a concept from the software engineering domain, describing symptoms which are likely to produce errors in execution and maintenance. By adapting the concept of code smells to the spreadsheet domain, it is possible to generate risk maps visualizing code smells and thus to indicate the exposure to risk at certain locations within a spreadsheet. Again, the objective of the present paper is to provide a formal foundation for the definition of those code smell metrics.

While the present paper seeks for a conceptual model for measuring spreadsheet complexity and assessing spreadsheet risk, Bregar [Bregar, 2004] defines complexity metric based on a mathematical model. He proposes a set of metrics, which are mostly adapted from the domain of software engineering, and provides a mathematical definition for most of them. However, while this type of formal definition enables a reliable reproducibility, an underlying conceptual model for spreadsheet complexity – as proposed by the present work – explicitly outlines the aspects of spreadsheets, which are considered to be drivers for the complexity and risk. Hence, contrary to the mathematical model by Bregar [Bregar, 2004], the conceptual model facilitates the identification and derivation of additional metrics from other domains, e.g., linguistics (as described in Section 5 of this paper).

Similarly, Hodnigg and Mittermeir [Hodnigg and Mittermeir, 2008] formally define complexity metrics based on a mathematical and graph-based notation, whereas the graph describes the dependencies between cells determined by cell references within formulas. The purpose of those metrics is to visualize the conceptual model of the spreadsheet in order to facilitate a spreadsheet's maintenance. They also distinguish between two types of metrics, namely general indicators capturing data and structure-related aspects of the spreadsheet, and formula complexity metrics which basically relate to the metrics proposed by Hermans et al. [Hermans et al., 2012b] and Bregar [Bregar, 2004]. And again, while they provide formal definitions of metrics for measuring the complexity of spreadsheets, the purpose of the present paper's conceptual spreadsheet model is to explicitly capture the complexity drivers of spreadsheets. Hence this model can serve as a foundation for the works by Hodnigg and Mittermeir [Hodnigg and Mittermeir, 2008] as well as Bregar [Bregar, 2004].

Cunha et al. [Cunha, Fernandes, Peixoto and Saraiva, 2012] present a quality model for spreadsheets based on a common software engineering standard. Thereby they define metrics for quality aspects including functionality, maintainability, and usability. While many of those metrics are covered by the works of Hodnigg and Mittermeir [Hodnigg and Mittermeir, 2008] as well as Bregar [Bregar, 2004], the usability-related metrics by Cunha et al. also capture visual attributes (e.g., color-coding for different cell types) as aspects affecting the understandability and thus the perceived complexity of spreadsheets. Therefore, this work provides additional input for the derivation of the conceptual model as described in Section 4.

Seila and Banks [Seila and Banks, 1990] propose simulation as an alternative approach for assessing the risk of errors in spreadsheets. Thereby they simulate an uncertainty in input variables, and thus observe the dynamics of the investigated spreadsheet in particular. The simulation reveals how vulnerable the spreadsheet's output in case of erroneous inputs. However, the conceptual model as described later on in the present work does not capture the dynamics of spreadsheets, but – similar to Hermans et al. [Hermans et al., 2012b], Bregar [Bregar, 2004], Hodnigg and Mittermeir [Hodnigg and Mittermeir, 2008], and Cunha et al. [Cunha et al., 2012] – focuses on the structural aspects of spreadsheets.

Shubbak and Thorne [2016] present a software tool for identifying and assessing the risk of a given spreadsheet by taking into account the spreadsheet's nature, importance, use, and complexity. While they already use some basic complexity measures, their work can profit directly from the model as proposed by the paper at hand, since it enables a structured and model-based approach to complexity metrics.

**3 RESEARCH METHOD**

For deriving the conceptual model for spreadsheet complexity as described later on in Section 4, we applied the design-science research method as defined by Hevner et al. [Hevner, March, Park and Ram, 2004], and as depicted in Figure 1.

The environmental part of the Information Systems Research Framework [Hevner et al., 2004] was already introduced in Section 1 by motivating the need for risk assessment in spreadsheets and a structured approach to the same. Furthermore, we design the conceptual model for spreadsheet complexity on the foundations of the existing knowledge base (see Section 2) which already describe different indicators and aspects of complexity in spreadsheets. In this sense, the model as presented in this work integrates the findings of the related papers and proposes a comprehensive conceptual model capturing all aspects, which were identified as potential drivers for complexity in spreadsheets. The conceptual model for spreadsheet complexity forms the artifact as defined by the Information Systems Research Framework. Its design is affected by both the relevance of the environmental part and the rigor through the knowledge base [Hevner et al., 2004]. We use an analytical and descriptive evaluation method to justify and argument for the conceptual model's utility and correctness. For this, we apply the already proposed spreadsheet complexity metrics on the one hand, and apply them to two common spreadsheet corpora as described in Section 6.

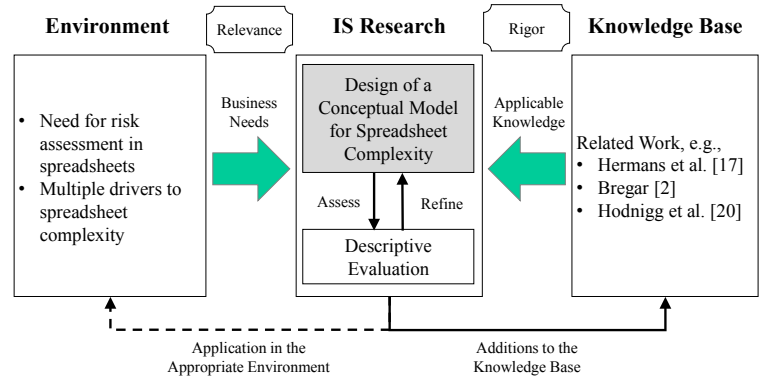

Figure 1: Overview over this work's research methodology based on the Information Systems Research Framework by Hevner et al. [Hevner et al., 2004].

In addition to the design-science research method, we also follow a quantitative descriptive content analysis method [Neuendorf, 2002] – in particular in the evaluation phase of the Information Systems Research Framework. Thereby, the quantitative metrics as described in Section 5 indicating formula complexity were derived from the scientific discipline of linguistics. Consequently, we methodologically transfer established insights from this discipline to determine metrics representing complexity of spreadsheets based on the proposed conceptual model for measuring spreadsheet complexity.

**4 A CONCEPTUAL MODEL FOR SPREADSHEET COMPLEXITY**

Based on the design-science research method as well as the related work about measuring and assessing the complexity of spreadsheets, we propose the conceptual model in Figure 2, capturing and integrating all aspects of spreadsheets, which were already identified as potential drivers for complexity.

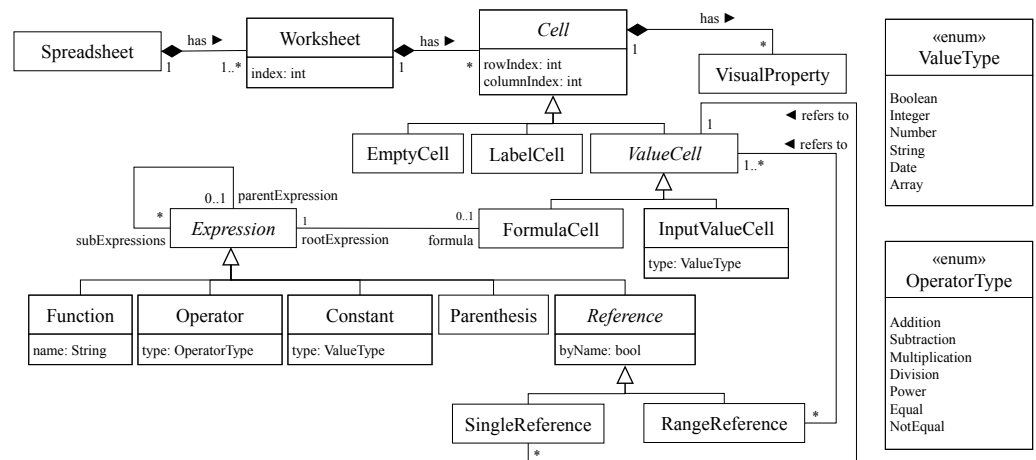

Figure 2: A conceptual and integrated model for measuring complexity in spreadsheets.

The concept of worksheets not only forms a basic part of a spreadsheet's structure, but is also an important aspect in the context spreadsheet complexity. For example,

Hermans et al. [Hermans, Pinzger and van Deursen, 2012a] use the worksheet information for determining interdependencies between them. A high coupling between worksheets indicates a code smell and thus implies an increased complexity.

A worksheet consists of cells that are organized in a two-dimensional grid. As shown by Rajalingham et al. [Rajalingham, Chadwick, Knight and Edwards, 2000], the placement of formulas and location of cells in general has a huge impact on the perceived complexity and thus understandability of spreadsheets. Therefore, we include the coordinate of the cell within a worksheet as an attribute of the *Cell* class. As suggested by Cunha et al.[Cunha et al., 2012], a cell's visual properties, e.g., its color, contribute to a spreadsheets understandability and perceived complexity, wherefore we add the class *VisualProperty* to the model. Due to a lack of evidence from related literature we do not specify this concept in detail, e.g., by defining which kind of visual properties are affecting the understandability of a spreadsheet (apart from the already mentioned color of cells [Cunha et al., 2012]).

Hodnigg et al. [Hodnigg and Mittermeir, 2008] differentiate between different kind of cells, namely empty cells, label cells, and value cells. Empty cells do not even contain any data or text. Label cells are not referenced by formulas and only serve as documentation and description of value cells, which in turn contain the actual data of the spreadsheet. Value cells can be further classified into cells containing formulas and cells serving as input fields for the spreadsheet data. Thereby, the *InputValueCell* class captures the type of its data, which is also an aspect potentially affecting the spreadsheet's complexity [Cunha et al., 2012]. In this context, the *ValueType* enumeration containing possible value types strongly depends on the actual spreadsheet software and has to be adapted accordingly when applying the model to a certain spreadsheet tool. The class *FormulaCell* represents cells which compute their value based on an expression – the formula. Hermans et al. [Hermans et al., 2012b], Bregar [Bregar, 2004], and Hodnigg et al. [Hodnigg and Mittermeir, 2008] identify the level of nesting of formulas – with respect to the formula's representation as an abstract syntax tree (AST) – as a complexity driver for spreadsheets. This AST is realized in the model in Figure 2 by the *Expression* class and its reflexive association. Thereby the formula refers to exactly one expression, which forms the root of a potentially nested tree structure of different kinds of expressions.

The number of conditionals (or decision count) within a formula is also considered to be a measure for its complexity [Bregar, 2004]. The occurrence of functions like *lookup* and *offset* adds further complexity to the formula [Hodnigg and Mittermeir, 2008]. We define the *Function* class as a concrete expression to capture those different kinds of functions in spreadsheet formulas. Furthermore, we differentiate between *Operator* (e.g., *plus* and *minus* for the arithmetic addition and subtraction), *Constant* of different types, and *Parenthesis* expressions in order to capture the diversity and nestedness of formulas. Again, the set of operator types and the set of value types strongly depends on the actual spreadsheet software.

As suggested by Hermans et al. [Hermans et al., 2012b], Bregar [Bregar, 2004], and Hodnigg et al. [Hodnigg and Mittermeir, 2008], the dependencies between formulas are one of the main drivers for spreadsheet complexity. Therefore, the conceptual model as proposed in this work captures those dependencies by the *Reference* and *Range* classes. The former one describes references to single cells, while the latter one represents a reference to a one- or two-dimensional cell block. According to Hermans et al. [Hermans et al., 2012b], range references have an even higher impact to a spreadsheet's complexity than references to a single cell. Furthermore, in another work [Hermans et al., 2012a] they suggest that cell references by name have to be differentiated from references by grid coordinates. For this reason, we add the Boolean attribute *byName* to the *Reference* class.

The integrated conceptual model in Figure 2 describes a network of cells, whereas most of its nodes contain an abstract syntax tree representing a formula. This model not only integrates knowledge about aspects for spreadsheet complexity based on existing complexity metrics, but also serves as a starting point for identifying new ways and metrics for quantifying complexity in related domains

**5 APPLICATION OF THE MODEL TO DEFINE COMPLEXITY METRICS**

Measures of complexity are well known in various scientific disciplines to get a theoretical and empirical insight into a system and its behavior. However, our research focuses on two disciplines that investigate different artifacts, namely software engineering and linguistics. Both deal – at least to some extent – with the analysis of man-made objects, which is software on the one hand, and language, i.e. text, on the other hand. The following two sections summarize the integration of the two disciplines regarding their understanding and measurement of complexity and show how these fit to our conceptual model of spreadsheet complexity.

**5.1 Metrics in Software Engineering**

Over the years, software engineering (SE) has become a mature scientific discipline dealing with a variety of challenges, i.e. planning, development, maintenance, etc. Metrics in SE are used by industry and research to understand and improve both software products and software development processes. Various metrics exist covering different aspects of SE processes and artifacts. However, validation of such metrics is difficult to generalize and mostly done ad-hoc [Meneely, Smith and Williams, 2013]. Therefore, we refer to established SE metrics, which were already adopted to the area of spreadsheets by Hermans et al. [Hermans et al., 2012b], Bregar [Bregar, 2004], Hodnigg et al. [Hodnigg and Mittermeir, 2008], and Cunha et al. [Cunha et al., 2012]. We show that existing and well-studied metrics can be redefined through our model (see Table 1). We exemplarily select 16 representative metrics covering all aspects of our model, except VisualProperty class, since there is no respective concrete metric definition in related literature.

**5.2 Metrics in Linguistics**

As mentioned in the introduction, linguists have eagerly defined metrics and indicators representing linguistics features on various levels. Graesser and McNamara et al. have defined over 100 different indices for evaluation of text and discourse [Graesser, McNamara, Louwerse and Cai, 2004; McNamara, Graesser, McCarthy and Cai, 2014]. Thereby they classified their metrics regarding the complexity and difficulty to which they contribute. This classification covers categories such as "Descriptive", "Readability", "Referential Cohesion", "Lexical Diversity", "Syntactic Complexity", etc. [McNamara et al., 2014]. This classification also contributes to a common understanding of a multilevel framework for discourse comprehension [Graesser and McNamara, 2011]. The levels cover the surface code, textbase, situation model, rhetorical structure, and pragmatic communication level. This structure refers to the rationale, that distinct linguistic properties influence the text on a different level, such as different entities in spreadsheets contribute to its complexity differently. However, the two top-most levels, namely rhetorical structure and pragmatic communication level cannot easily be transferred to the domain of spreadsheet complexity. Surface code, textbase, and situation model on contrary cover technical and structural properties of text and can be reused for analysis of spreadsheet formulas based on a constructive model. Based on the metrics provided by Grasser and McNamara et al. [McNamara et al., 2014] we selected a subset for the analysis and adapted those to the domain of spreadsheet formulas (see Table 2).

Table 1. A selection of complexity metrics from the software engineering domain.

| Name | Description |
| --- | --- |
| Average AST depth per formula | The average depth of the abstract syntax tree which is formed by the *Expressions* of a *FormulaCell*. |
| Max AST depth per formula | |
| Number of formula cells | The size of the spreadsheet which is determined by the number/ratio of different kind of *Cells*, e.g., *ValueCell* and *FormulaCell*. |
| Ratio of formula cells to non-empty cells | |
| Number of input cells | |
| Ratio of input cells to non-empty cells | |
| Ratio of formula cells to input cells | |
| Number of distinct formulas | Number of distinct *FormulaCells* which have an equally structured AST which only differs with respect to its *References*. |
| Average fan-out per formula | Number of incoming and outgoing references of a *FormulaCells*. |
| Max fan-out per formula | |
| Average fan-in per formula | |
| Max fan-in per formula | |
| Average number of conditionals per formula | Number of Boolean expressions within one *FormulaCell*, e.g., functions with the names *IF*, *COUNTIF*, *SUMIF*, etc. |
| Max number of conditionals per formula | |
| Average spreading factor per formula | Maximal Euclidian distance between *References* in a *Formula*. In this sense, *rowIndex* and *columnIndex* of the *Cell* and *index* of the *Worksheet* represent the x, y, z coordinates of a three-dimensional space. |
| Max spreading factor per formula | |

Table 2. Selected metrics from the linguistics domain.

| Name | Description |
| --- | --- |
| Average number of functions per formula | Capture the lexical diversity of a *FormulaCell* by counting the number of used *Functions*. |
| Max number of functions per formula | |
| Average number of distinct functions per formula | Capture the lexical diversity of a *FormulaCell* by counting the number of used *Functions*, without counting duplicates. |
| Max number of distinct functions per formula | |
| Average number of elements per formula | Capture the connective incidence of a *FormulaCells* by counting the number of expressions within its AST. |
| Max number of elements per formula | |

The selection of six linguistic metrics enables the quantification of properties of spreadsheet formulas. Additionally, the number of references in texts are well-known linguistic metrics. However, since they are already covered through Software Engineering metrics, they are omitted in this table. Our selection represents the most fundamental quantifications that can be used straightforward.

## 6 APPLICATION OF METRICS

Based on the conceptual model for spreadsheet complexity as described in Section 4, we applied the metrics from Section 5 to the EUSES and Enron spreadsheet corpora. The application of the metrics enables the comparison of them with respect to different aspects of spreadsheet complexity on the one hand, and the determination of correlations between potential drivers to complexity on the other hand. The EUSES spreadsheet corpus [Fisher and Rothermel, 2005] contains over 4,000 spreadsheets which were mainly crawled with search engines. The Enron corpus [Hermans and Murphy-Hill, 2014] consists of over 15,000 industrial spreadsheets gathered from the Enron Corporation. Hermans et al. [Hermans and Murphy-Hill, 2014] already state the spreadsheets of the Enron corpus are significantly more smelly and thus they are considered to be more complex than those from the EUSES spreadsheet corpus. This claim can be partially supported by the statistics in Table 3, in particular by the metrics from the domain of SE. For example, the values for the fan-in and fan-out metrics capturing the incoming and outgoing references of formula cells are much higher for the spreadsheets of the Enron corpus than for those of the EUSES corpus. Interestingly enough, though, there is no significant difference between spreadsheets of those two corpora from a linguistics perspective.

Table 3. Average complexity values for the EUSES and Enron spreadsheet corpora

|  | EUSES | Enron |
|---|---|---|
| Ratio of spreadsheets with formulas | 43 % | 58 % |
| *Software Engineering Metrics* | | |
| Average AST depth per formula | 1.92 | 1.51 |
| Max AST depth per formula | 4.63 | 4.47 |
| Number of formula cells | 350 | 2107.53 |
| Ratio of formula cells to non-empty cells | 0.23 | 0.30 |
| Number of input cells | 4931.90 | 11170.50 |
| Ratio of input cells to non-empty cells | 1.55 | 5.38 |
| Ratio of formula cells to input cells | 3.63 | 2.54 |
| Number of distinct formulas | 3.13 | 10.50 |
| Average fan-out per formula | 167.94 | 473.27 |
| Max fan-out per formula | 476.79 | 4709.88 |
| Average fan-in per formula | 0.93 | 7.70 |
| Max fan-in per formula | 9.20 | 50.53 |
| Average number of conditionals per formula | 0.09 | 0.07 |
| Max number of conditionals per formula | 0.27 | 0.33 |
| Average spreading factor per formula | 148.13 | 374.80 |
| Max spreading factor per formula | 350.94 | 1522.60 |
| *Linguistics Metrics* | | |
| Average number of functions per formula | 0.53 | 0.48 |
| Max number of functions per formula | 1.00 | 1.17 |
| Average number of distinct functions per formula | 0.51 | 0.47 |
| Max number of distinct functions per formula | 0.88 | 0.93 |
| Average number of elements per formula | 5.43 | 4.25 |
| Max number of elements per formula | 14.64 | 15.71 |

As shown by the histogram in Figure 3, the *ratio of formula cells* metric indicates that most spreadsheets have a low complexity. However, at the same time there is also a considerable amount of spreadsheets having high values for this metric wherefore they are considered to be very complex. This distribution of the complexity is even more significant for the *average fan-out metric* in Figure 4.

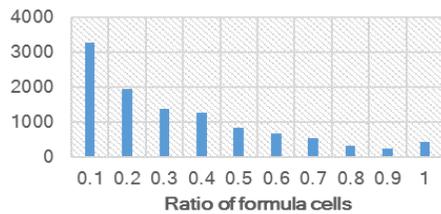 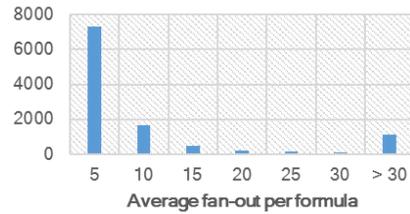

Figure 3. A histogram showing the number of spreadsheets by the ratio of formula cells.

Figure 4. A histogram showing the number of spreadsheets by the average fan-out per formula.

Computing the correlation coefficient for each pair of metrics based on the almost 20,000 spreadsheets of both corpora reveals that only those metrics correlate to each other which capture the same concepts of the conceptual model in Figure 2 (e.g., number of input cells and average fan-out). This is interesting in the sense that the aspects captured by the conceptual model for spreadsheet complexity are mostly independent from each other, and a high complexity with respect to a certain aspect does not imply a high complexity with respect to another one.

## 7 CONCLUSION

This paper presents an approach that fits neatly into the research dealing with the analysis of spreadsheets complexity. It proposes a conceptual and integrated model of spreadsheets allowing a detailed investigation of today's spreadsheets, which is the answer to the first research question from Section 1. Beyond the qualitative structuring of attributes and properties of spreadsheet formulas, the paper argues for a quantification of those. Thereby, it transfers well-studied and established results from two different but related domains dealing with the determination of complexity and understandability of man-made artifacts, namely software engineering and linguistics. In both disciplines, metrics to evaluate complexity are common and have been repeatedly reported as useful. Our paper maps those metrics to our model in a way that relevant aspects of spreadsheet complexity are covered by at least one metric. Thereby, we answer the second research question raised in the introduction. We evaluate these metrics by their application to a large set of existing spreadsheets, namely the EUSES and the Enron corpus. The empirical evaluation shows that many spreadsheets used in industry do not even contain formulas. However, if they do, the distribution of the *ratio of formula cells* metric indicates that most have a supposedly low complexity, whereas there are also numerous spreadsheets with an increased complexity. The *average fan-out* distribution shows a similar distribution: While most spreadsheets have only a *little average fan out* (<5), there is also a considerable amount of spreadsheets with a very high value for this metric (>30).

Ongoing research may integrate temporal aspects into the conceptual model for spreadsheet complexity, e.g., the temporal evolution of complexity metrics, but also the impact of changes of a spreadsheet to its complexity. Thereby, the model would be able to not only capture static and structural aspects of spreadsheets, but also the respective dynamics. Apart from this, the proposed model does not capture macros which usually have a high impact on the understandability and complexity of spreadsheets. This would be an additional aspect which should be added to a holistic complexity model of spreadsheets. Furthermore, the proposed conceptual model can serve as a foundation for deriving new complexity measures through adoption from other scientific domains. For example, due to the network structure of formulas as defined by the proposed conceptual model, sophisticated graph algorithms could be applied in order to reveal structural aspects. Similarly, methods from social network analysis could be adapted to the field of spreadsheets. Another example for metrics which can be adapted to the domain of spreadsheets are measures describing the diversity of elements within a certain context by applying the Shannon entropy, e.g., the entropy of functions or operators within a function. As these examples suggest, our conceptual model for spreadsheet formulas can serve as a theoretical foundation and starting point for upcoming metrics, measurements, and analysis in general.

## 8 REFERENCES

Bradley, L. and K. McDaid (2009) "Using Bayesian Statistical Methods to Determine the Level of Error in Large Spreadsheets", *Proceedings of the International Conference on Software Engineering*, pp. 351–354.

Bregar, A. (2004) "Complexity Metrics for Spreadsheet Models", *Proceedings of the European Spreadsheet Risks Information Group*, pp. 85–93.

Bretz, J. (2009) "OpRisk individuelle Datenverarbeitung. Management der operationellen Risiken aus dem Einsatz individueller Datenverarbeitung (IDV)", *Bank-Praktiker rechtssicher, revisionsfest, risikogerecht*, (6), pp. 294–298.

Bretz, J. (2012) *Prüfung IT im Fokus von MaRisk und Bundesbank: Verstärkter IT-Fokus in Sonderprüfungen*, Heidelberg: Finanz Colloquium Heidelberg.

Caulkins, J. P., E. L. Morrison, T. Weidemann (2007) "Spreadsheet Errors and Decision Making: Evidence from Field Interviews", *Journal of Organizational and End User Computing*, (19)3, pp. 1–23.

Chan, Y. E. and V. C. Storey (1996) "The Use of Spreadsheets in Organizations: Determinants and Consequences", *Information & Management*, (31)3, pp. 119–134.

Cunha, J., J. P. Fernandes, C. Peixoto, J. Saraiva (2012) "A Quality Model for Spreadsheets", *Proceedings of the International Conference on the Quality of Information and Communications Technology*.

DuBay, W. H. (2004) *The Principles of Readability*: ERIC.

Fisher, M. and G. Rothermel (2005) "The EUSES Spreadsheet Corpus: A Shared Resource for Supporting Experimentation with Spreadsheet Dependability mechanisms", *Proceedings of the Workshop on End-User Software Engineering*, pp. 47–51.

Flesch, R. (1948) "A New Readability Yardstick", *Journal of Applied Psychology*.

Gable, G. G., C. M. Yap, M. N. Eng (1991) "Spreadsheet Investment, Criticality, and Control", *Proceedings of the Hawaii International Conference on System Sciences*, pp. 153–162.

Graesser, A. C. and D. S. McNamara (2011) "Computational Analyses of Multilevel Discourse Comprehension", *Topics in cognitive science*, (3)2, pp. 371–398.

Graesser, A. C., D. S. McNamara, M. M. Louwerse, Z. Cai (2004) "Coh-Metrix: Analysis of Text on Cohesion and Language", *Behavior Research Methods, Instruments, & Computers*, (36)2, pp. 193-202.

Hall, M. J. J. (1996) "A Risk and Control-Oriented Study of the Practices of Spreadsheet Application Developers", *Proceedings of the Hawaii International Conference on System Sciences*, pp. 364–373.
Hermans, F. and E. Murphy-Hill (2014) *Enron's Spreadsheets and Related Emails: A Dataset and Analysis*.

Hermans, F., M. Pinzger, A. van Deursen (2012a) "Detecting Code Smells in Spreadsheet Formulas", *Proceedings of the International Conference on Software Maintenance*.
Hermans, F., M. Pinzger, A. van Deursen (2012b) "Measuring Spreadsheet Formula Understandability", *Proceedings of the European Spreadsheet Risks Information Group*, pp. 77–96.

Hermans, F., M. Pinzger, A. van Deursen (2014) "Detecting and Refactoring Code Smells in Spreadsheet Formulas", *Empirical Software Engineering*, pp. 1–27.

Hevner, A. R., S. T. March, J. Park, S. Ram (2004) "Design Science in Information Systems Research", *Management Information Systems Quarterly*, (28)1, pp. 75–105.

Hodnigg, K. and R. T. Mittermeir (2008) "Metrics-Based Spreadsheet Visualization: Support for Focused Maintenance", *Proceedings of the European Spreadsheet Risks Information Group*.

Janvrin, D. and J. Morrison (2000) "Using a Structured Design Approach to Reduce Risks in End User Spreadsheet Development", *Information & Management*, (37)1, pp. 1–12.

Köhler, R. (2005) *Quantitative Linguistik*, Berlin [u.a.]: De Gruyter.

McNamara, D. S., A. C. Graesser, P. M. McCarthy, and Z. Cai (2014) *Automated Evaluation of Text and Discourse with Coh-Metrix*: Cambridge University Press.

Meneely, A., B. Smith, L. Williams (2013) "Validating Software Metrics: A Spectrum of Philosophies", *ACM Trans. Softw. Eng. Methodol.*, (21)4, pp. 24:1-24:28.

Neuendorf, K. A. (2002) *The Content Analysis Guidebook*: Sage.

Panko, R. R. (2000) "Spreadsheet Errors: What We Know. What We Think We Can Do", *Proceedings of the European Spreadsheet Risks Information Group*, pp. 7–17.

Panko, R. R. (2006) "Facing the Problem of Spreadsheet Errors", *Decision Line*, (37)5, pp. 8–10.

Panko, R. R. and D. N. Port (2012) "End User Computing: The Dark Matter (and Dark Energy) of Corporate IT", *Journal of Organizational and End User Computing*, pp. 4603–4612.

Pemberton, J. D. and A. J. Robson (2000) "Spreadsheets in Business", *Industrial Management & Data Systems*, (100)8, pp. 379–388.

Powell, S. G., K. R. Baker, B. Lawson (2008) "A Critical Review of the Literature on Spreadsheet Errors", *Decision Support Systems*, (46)1, pp. 128–138.

Powell, S. G., K. R. Baker, B. Lawson (2009) "Impact of Errors in Operational Spreadsheets", *Decision Support Systems*, (47)2, pp. 126–132.

Rajalingham, K., D. Chadwick, B. Knight, D. Edwards (2000) "Quality Control in Spreadsheets: A Software Engineering-Based Approach to Spreadsheet Development", *Proceedings of the Hawaii International Conference on System Sciences*.

Reschenhofer, T. and F. Matthes (2015) "A Framework for the Identification of Spreadsheet Usage Patterns", *Proceedings of the European Conference on Information Systems*.

Scaffidi, C., M. Shaw, B. Myers (2005) "Estimating the Numbers of End Users and End User Programmers", *Proceedings of the Symposium on Visual Languages and Human-Centric Computing*, pp. 207–214.

Seila, A. F. and J. Banks (1990) "Spreadsheet Risk Analysis Using Simulation", *Simulation*, (55)3, pp. 163–170.

Shubbak, M. H. and S. Thorne (2016) "Development and Experimentation of a Software Tool for Identifying High Risk Spreadsheets for Auditing", *Proceedings of the European Spreadsheet Risks Information Group*.

Teo, T. S. H. and J. E. Lee-Partridge (2001) "Effects of Error Factors and Prior Incremental Practice on Spreadsheet Error Detection: An Experimental Study", *Omega*, (29)5, pp. 445–456.

Waltl, B. and F. Matthes (2014) "Towards Measures of Complexity: Applying Structural and Linguistic Metrics to German Laws", *Jurix 2014: Legal Knowledge and Information Systems*.